# Financial Structure, Firm Size and Financial Growth of Non-Financial Firms Listed at the Nairobi Securities Exchange

## David Haritone Shikumo, Oluoch Oluoch, Joshua Matanda Wepukhulu

School of Business, Jomo Kenyatta University of Agriculture and Technology, Nairobi, Kenya

**Email address:**
dahikumo@gmail.com (David Haritone Shikumo)

**To cite this article:**
David Haritone Shikumo, Oluoch Oluoch, Joshua Matanda Wepukhulu. Financial Structure, Firm Size and Financial Growth of Non-Financial Firms Listed at the Nairobi Securities Exchange. *Journal of Finance and Accounting*. Vol. 11, No. 1, 2023, pp. 12-25.
doi: 10.11648/j.jfa.20231101.12

**Received**: December 23, 2022; **Accepted**: January 16, 2023; **Published**: January 30, 2023

**Abstract:** A significant number of the non-financial firms listed at the Nairobi Securities Exchange (NSE) have been experiencing declining financial performance and financial growth, which deter investors from investing in such firms. Hence, the study aimed at establishing the effect of financial structure on the financial growth of non-financial firms listed at the NSE. The study was guided by the Modigliani-Miller theory, Agency Theory, Pecking Order Theory, Trade-off Theory, Market Timing Theory, and Theory of Growth of the Firm. An explanatory research design was adopted. The study's target population comprised 45 non-financial firms listed at the NSE for a period of ten years, from 2008 to 2017. The panel model revealed that short-term debt, long-term debt, retained earnings, and share capital explain 61.36% of variations in financial growth as measured by growth in earnings per share and 65.57% of variations in financial growth as measured by growth in market capitalization. Short-term debt has a positive and significant effect on financial growth as measured by growth in earnings per share ($\beta=0.024095$, $p=0.013$) and growth in market capitalization ($\beta=0.028529$, $p=0.006$). Long-term debt has a positive and significant effect on financial growth as measured by growth in earnings per share ($\beta=0.864088$, $p=0.000$) and growth in market capitalization ($\beta=0.958656$, $p=0.000$). Retained earnings have a positive and significant effect on financial growth as measured by growth in earnings per share ($\beta=0.951749$, $p=0.015$) and growth in market capitalization ($\beta=0.043784$, $p=0.004$). Further, share capital has a positive and significant effect on financial growth as measured by growth in earnings per share ($\beta=0.007016$, $p=0.000$) and growth in market capitalization ($\beta=0.09635$, $p=0.001$). Firm size significantly intervenes in the effect of financial structure on the financial growth of non-financial firms listed at the NSE measured using growth in earnings per share. However, Firm size does not intervene in the effect of financial structure on the financial growth of non-financial firms listed at the NSE measured using growth in market capitalization. The study concludes that short-term debt, long-term debt, retained earnings, and share capital positively influence financial growth as measured by both growth in earnings per share and growth in market capitalization. The study recommends that the management of non-financial firms listed at the NSE to balance financing a firm using debt and equity. The study also recommends that the management of non-financial firms listed at NSE to encourage its shareholders to re-invest back their earnings rather than consuming them as dividends. Financial structure varies significantly depending on the sector in which a firm operates. There is a need to conduct a comparison study to establish the effect of financial structure on the financial growth of non-financial firms versus financial firms listed at the NSE. The comparison study will tell which form of financing is appropriate for non-financial firms and which one is appropriate for financial firms.

# 1. Introduction

## 1.1. Background of the Study

Financial structure is the way a firm finances its assets and operations through some combination of debt and equity that a firm deems appropriate to enhance its operations [71]. The determination of a firm's optimal financial structure is vital in deciding how much money should be borrowed and the best mixture of debt and equity to fund business operations [83]. Therefore, the choice among an ideal proportion of debt and equity can affect the firm's value and its financial growth. The choice of the appropriate mix of different sources of financing such as short-term debt, long-term debt, retained earnings, and share capital is one of the critical decisions that need to be taken by the managers of the firm [47]. A firm can also arrange lease financing, use warrants, issue convertible bonds, sign forward contracts, or trade bond swaps. Firms can also issue dozens of distinct securities in countless combinations to maximize overall market value [47]. Financial structure is therefore very critical and fundamental in the business life cycle, not only to maximize shareholders' wealth but also due to its impact on the financial growth of the firm [39].

Financial structure is the result of some decisions that managers take in order to support long-term investments and identify appropriate sources of financing to contribute to the optimal development of the firm [75]. During the financial crisis, firms that are vulnerable to the shocks in the financial markets absorb the negative impact earlier than other firms. Hence, financial difficulties can lead to bankruptcy and can be the result of wrong decisions in choosing the financial structure [30]. The study of the financial structure has been for a long time, the central theme of concern in corporate firms since the theorem of Modigliani and Miller came to rule in finance literature discussing all the inapplicability of the financial structure for real decisions [52].

Financial growth is a measure of the efficient utilization of assets by a firm from principal business mode to generate revenues [2]. Financial growth is a general measure of the overall financial health of a firm over a given period [69]. According to Buvanendra, S., Sridharan, P., and Thiyagarajan, S., the financial growth of a firm is measured as the growth of market capitalization [21]. Market capitalization refers to the total dollar market value of a company's outstanding shares [79]. Market capitalization is calculated by multiplying a company's shares outstanding by the current market price of one share [22]. The return on investment, return on assets, market value, and accounting profitability reflect the financial growth of firms [70].

## 1.2. Statement of the Problem

Financial structure decision is a very critical decision with great implications for the firm's financial growth [8]. In order to maximize the firm's financial growth, the management must carefully consider the financial structure decision [30]. However, financing decisions are complex and vary between the firms. For instance, if financing is made using the wrong combination of debt and equity, the performance and financial growth of the firm will be negatively affected [86]. In addition, despite the substantial theoretical developments over the past several decades, the gap between financial structure theories and practices still exists [85]. For instance, the Modigliani and Miller theory demonstrated the irrelevance of capital structure in a firm's value while the pecking order, agency, and trade-off theories demonstrate that financial structure affects a firm's value as well as financial growth ([52, 60]). These theories have always been a subject of considerable debate due to inherent controversies. The extensive debate from diverse perspectives has made this issue more complex.

The financial growth of a significant number of non-financial firms listed at the Nairobi Securities Exchange has been declining [56]. For instance, the financial growth of non-financial firms listed at NSE dropped from 4.2% in 2016 to 3.7% in 2017 [63]. In addition, more than 56% of non-financial firms quoted at NSE recorded a deteriorating market capitalization trend between 2011 and 2015 [87]. The decline in financial growth deters lenders from lending to such firms [61]. It is not clear whether an optimal financial structure can lead to the financial growth of the firm. In addition, there exists a great dilemma for scholars, business managers, and investors among other stakeholders as to whether there exists an optimal financial structure that can maximize the financial growth of the firm.

The majority of the studies conducted; ([5, 56, 83, 35, 22]) focused on the capital structure while basing their argument on the accounting concept. Unlike financial structure, short-term liabilities do not contribute to capital structure ([71, 56]). There are also inconsistencies in results from previous empirical studies on the effects of long-term debt and short-term debt on the financial performance of listed non-financial firms ([71, 56]). Based on the reviewed studies, the effect of financial structure on the financial growth of non-financial firms listed at the Nairobi Securities Exchange is under-researched. For instance, Pouraghajan, A., Malekian, E., Emamgholipour, M., Lotfollahpour, V., and Bagheri, M. M. conducted a study on the relationship between capital structure and firms' performance for listed firms on the Tehran Stock Exchange [75]. Bokhari, H. W., and Khan, M. A. determined the impact of capital structure on firm's performance of non-financial firms in Pakistan [19]. Opungu, J. A. conducted a study to investigate the effect of capital structure on the profitability of non-financial firms listed on the Nairobi Stock Exchange (NSE) [71]. Mohammadzadeh, M., Rahimi, F., Rahimi, F., Aarabi, S. M., and Salamzadeh, J. conducted a study on how capital structure affects the profitability of pharmaceutical firms in Iran [54].

Moreover, Salawu, R. O., and Agboola, A. A. examined the determinants of capital structure of large non-financial firms in Nigeria using a panel of thirty-three (33) large firms [78]. Ater, D. K. conducted a study on capital structure and the value of non-financial firms listed at the Nairobi Securities Exchange [10]. In addition, Githire, C., and Muturi,

W. conducted a study to examine the effect of capital structure on the performance of firms listed at the NSE [34]. El-Chaarani, H. conducted a study on the impact of financial structure on the performance of European listed firms [27]. Oladele, S. A., Omotosho, O., and Adeniyi, S. D. conducted a study on the effect of capital structure on the performance of Nigerian-listed manufacturing firms from 2004-2013 [65]. Tsoy, L., and Heshmati, A. conducted a study on the impact of financial crises on the dynamics of the capital structure of listed non-financial firms in Korea [84].

Thus, a review of the literature indicates that the majority of past empirical studies ([68, 5, 89, 65, 40, 18, 26, 33, 75, 35, 19, 12, 22, 71, 10, 34, 56]) analyzed the effect of financial structure on the financial performance of the firms. The preceding scholars have only illustrated the theoretical understanding of the effect of financial structure on performance. None of the reviewed studies precisely examined the effect of short-term debt, long-term debt, retained earnings, and share capital on the financial growth of non-financial firms listed at the Nairobi Securities Exchange with an intervening effect of the firm size. The study intended to fill this knowledge gap by focusing on the effect of financial structure on the financial growth of non-financial firms listed at the Nairobi Securities Exchange.

### 1.3. Research Objectives

1) To determine the effect of financial structure on the financial growth of non-financial firms listed at the Nairobi Securities Exchange.
2) To examine the intervening effect of firm size on the effect of financial structure on the financial growth of non-financial firms listed at the Nairobi Securities Exchange.

## 2. Literature Review

### 2.1. Theoretical Review

The study is guided by the Modigliani-Miller Theory, Agency Theory, Pecking Order Theory, Trade-off Theory, Market Timing Theory, and Theory of Growth of the Firm.

#### 2.1.1. The Modigliani-Miller Theorem (MM Theory)

Modigliani, F. and Miller, M. advanced the capital structure irrelevance theory [52]. The Modigliani-Miller theorem on the irrelevancy of capital structure implicitly assumes that the market possesses full information about the activities of firms and that the information asymmetry influences financial growth [51]. Modigliani, F. and Miller, M. assert that when the market conditions are perfect, the value of a firm's stocks is not determined by capital structure decisions [52]. The MM capital structure irrelevance theory presupposes that the capital mix is unrelated to the value of the firm [31]. This suggests that the valuation of a firm is irrelevant to the capital structure of a company. Whether a firm is highly leveraged or has a lower debt component has no bearing on its market value [6]. Rather, the market value of a firm is solely dependent on the operating profits of the company. Whether a firm is highly leveraged or has a lower debt component in the financing mix has no bearing on its value of a firm [1].

The Modigliani and Miller approach to capital theory, advocates the capital structure irrelevancy theory. Modigliani and Miller advocate capital structure irrelevancy theory, which suggests that the valuation of a firm is irrelevant to the capital structure of a company [4]. Moreover, the Modigliani - Miller proposition is based on the assumptions of a perfect capital market in which there are no transaction costs, no information asymmetry (investors have the same information as management about the firm's future investment opportunities), no bankruptcy costs (debt is risk-free regardless of the amount used), so no firm goes bankrupt, no taxes (no taxes exist either on individuals or companies) and investors can borrow at the same rate as corporations [91]. Further, management acts on the exclusive behalf of shareholders. These assumptions can be criticized on the grounds that imperfections in capital markets do exist, suggesting that different sources of financing may be relevant to the investment decision of the firm [91]. The study tested whether the mix of long-term debt and Short-term debt that non-financial firms apply in their financial structure influences their financial performance as well as financial growth.

#### 2.1.2. Agency Theory

Jensen, M. C., and Meckling, W. H. advanced the agency theory which states that a firm has an optimal financial structure that stimulates optimum financial growth [42]. The optimum financial structure is obtained by ensuring that agency costs that arise from the conflicts between the managers and owners of the business are reduced by having a certain proportion of debt in the financial structure [46]. The lowering of agency conflicts would lead to a reduction in agency costs which would lead to improved financial growth. The use of debt in the firm as observed by Jensen, M. C., and Meckling, W. H. can help to control and monitor managers in the firm to ensure that they follow objectives that are beneficial to the firm [42].

Agent theory captures the idea of agency costs that arise as a result of conflicts between managers, shareholders, and creditors. These conflicts are supposed to arise due to the inconsistency of interests [42]. Managers tend to use the firm's resources in projects that bring more personal benefits than maximizing the value of the firm. Shareholders can discourage such behavior through monitoring and control activities [36]. However, these actions also involve costs, called agency costs. Debt can reduce agency costs and affect the performance of the firm at the same time, by determining the managers to act in the interest of the firm rather than in their own interest [44]. Thus, the option of a firm to be financed through debt, reduces the cash flow available at the discretion of managers, reducing agency costs.

#### 2.1.3. Pecking Order Theory

Donaldson, G. [25] postulated this theory but it received its first rigorous theoretical foundation from Myers, S. C., and Majluf, N. S. [59]. Myers, S. C., and Majluf, N. S. assert that firms have a particular preference order for capital used to finance their business [59]. Pecking order theory predicts that due to the information asymmetry between the firm and outside investors regarding the real value of both current operations and future income streams and prospects, external capital will always be relatively costly compared to internal capital [66]. Myers, S. C., and Majluf, N. S. argued that if firms issue no new security but only use their retained earnings to support investment opportunities, the information asymmetry can be resolved [59]. This implies that issuing equity becomes more expensive as information asymmetry between insiders and outsiders increases hence leading to undervalued securities.

Managers will prefer financing new investments through internal sources (that is, retained earnings) first, if this source is not enough then managers seek external sources from debt second and equity last [81]. Thus, according to the pecking order theory firms that are profitable and, therefore, generate high earnings to be retained are expected to use less debt in their financial structure than those that do not generate high earnings, since they are able to finance their investment opportunities with retained earnings [28]. The pecking Order theory states that companies prioritize their sources of financing from internal financing to equity. Therefore, internal financing is used first then when that is depleted, debt is issued and when it is no longer sensible to issue any more debt, equity is issued [80].

### 2.1.4. Trade-Off Theory

The trade-off theory postulated by Myers, S. C. emphasizes a balance between tax savings arising from debt, a decrease in agent cost, and financial distress [58]. Trade-Off Theory claims that firms have the incentive to turn to debt as the generation of annual profits allows benefiting from the debt tax shields. Shahar, W. S. S., Shahar, W. S. S., Bahari, N. F., Ahmad, N. W., Fisal, S., and Rafdi, N. J. finds that the benefit of a tax shield is offset by the firm costs of financial distress and agency costs [82]. In other words, the optimal level of leverage is achieved by balancing the benefits from interest payments and the costs of issuing debt [41]. The balance between tax savings arising from debt, a decrease in agent cost, and financial distress has a significant effect on financial growth.

An important purpose of the theory is to explain the fact that, corporations are usually financed partly with debt and partly with equity. It states that there is an advantage to financing with debt, the tax benefits of debt and there is a cost of financing with debt, the costs of financial distress including bankruptcy costs of debt and non-bankruptcy costs. The marginal benefit of further increases in debt declines as debt increases, while the marginal cost increases, so a firm that is optimizing its overall value will focus on this trade-off when choosing how much debt and equity to use for financing [58].

### 2.1.5. Market Timing Theory

The market timing theory was proposed by Baker, M., and Wurgler, J. [15]. The theory postulates that financial structure is the cumulative outcome of past attempts to time the equity market [15]. The market timing theory, developed by Baker, M., and Wurgler, J. [15], starts from the idea that raising capital by issuing shares depends on market performance. In corporate finance, market timing involves in practice, issuing high-priced shares and repurchasing them at a lower price, in order to benefit from fluctuations in the ratio between the cost of equity and other forms of capital. The market timing theory assumes that firms time their equity issues whereby they will issue new stock when the stock price is perceived to be overvalued (high price) and repurchase their shares when there is undervaluation (low price) [49, 55]. This implies the firm's intent to take advantage of fluctuations in equity market valuations. As a result, fluctuations in stock prices will affect the firm's financial decision as well as its financial growth.

Two broad criticisms have been leveled at market timing theory. The first criticism, voiced by Alti, A. and Flannery, M. J., & Rangan, K. P., questions the longevity and overall economic significance of market timing [7, 29]. However, Huang, R., and Ritter, J. R. using aggregate measures of market valuation, find evidence of a long-lasting market timing effect on capital structure and Leary, M. T., and Roberts, M. R. find that shocks to equity valuation can persist for varying lengths of time ([38, 45]). The second criticism, as proposed by Hovakimian, A. contends that the negative relationship between market-to-book and leverage is not indicative of market timing [37]. Instead, the relationship is argued to be due to growth opportunities, which when high (low), lead firms to use more (less) equity financing. Hovakimian, A. also contends that the cross-sectional relationship between market-to-book and leverage dominates the temporal relationship [37]. Market timing theory was relevant to the study by enabling firms to decide whether to finance their investments via debt or equity. The firms either choose to finance through equity or debt based on the market situation.

### 2.1.6. Theory of Growth of the Firm

The theory was propagated by Penrose, E. T. [72]. Penrose argued that firms had no determinant to a long run or optimum size, but only a constraint on current period growth rates. According to the theory, financial means for expansion could be found through retained earnings, borrowing, and new issues of stock shares [72]. Retained earnings are one of the most important sources to finance new projects in emerging economies where capital markets are not well developed. However, firms in the start-up period, when an initial investment has not matured yet or whose investment projects are substantially larger than their current earnings, will not have enough financial means from retained earnings and will face a constraint in their growth project [74]. Firms in this situation may seek external sources of financing; however, the extent of borrowing could be limited by internal factors like high debt-equity ratios that would expose both borrowers and lenders to increased risk. In other cases, financing of growth projects may be limited by shallow financial markets. Rajan, R. G., and Zingales, L. found that industrial sectors with a great need for external finance grow substantially less in countries without well-developed financial markets [77].

However, the theoretical focus of growth of the firm theory prevents the development of a richer, more complex, contingency-based model of firm growth [73]. The growth of the firm theory does not consider the professionalization of management, the evolution of technology, and institutions that influence firms' growth. Penrose, E. T. preempts such criticisms by clearly emphasizing the central focus of the growth of the firm theory [72]. Another critique of the book is the testability of the growth of the firm theory. Unlike theoretical work today, which emphasizes constructs and relationships, Penrose mostly used case histories to develop some theoretical principles and logic, and acknowledged that testing them remained problematic. This theory is relevant to this study since it informs the dependent variable which is financial growth.

### 2.2. Conceptual Framework

A conceptual framework is a visual or written product, one that explains either graphically or in narrative form the main things to be studied, the key factors, concepts, or variables, and the presumed relationships among them [43]. Figure 1 shows the conceptual framework.

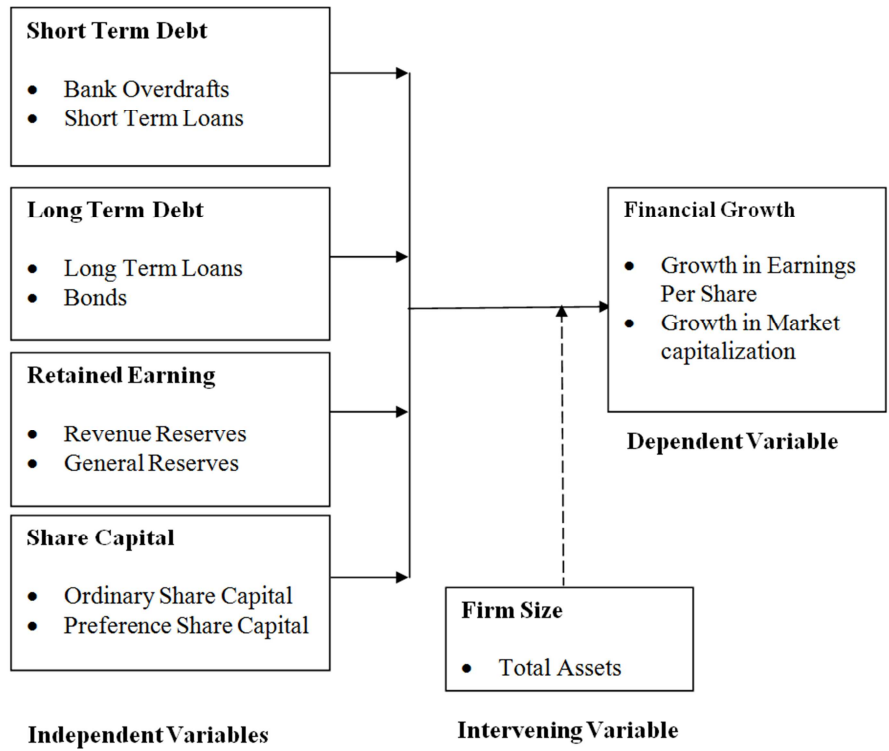

*Figure 1. Conceptual Framework.*

## 3. Research Methodology

The study adopted a positivist research philosophy. Positivism philosophy premises that, knowledge is based on facts and that no abstractions or subjective status of individuals are considered [20]. The study adopted an explanatory research design. The target population comprised 45 non-financial firms listed at the NSE for a period of ten years from 2008 to 2017. The study adopted a census technique where all non-financial firms listed at NSE were considered. The census approach is the total inclusion of all observations in the study [20]. The study used a secondary data collection template to collect the data. The data was analyzed using STATA 14 software. The study employed multiple regression under the panel data framework. Panel data regression was chosen for a number of reasons. First, panel data allowed for the control of individual heterogeneity, making it possible to exclude biases deriving from the existence of individual effects. Secondly, panel data yielded more informative data, more variability and less collinearity among variables, more degree of freedom, and more efficiency than cross-section data or time-series data. Thirdly, panel data was used to obtain consistent estimators in the presence of omitted variables. The panel model was: -

$$(Y)_{i,t} = \beta_0 + \sum_{t=1}^{\infty} \beta_i X_{i,t} + \varepsilon_{i,t} \qquad (1)$$

Where;
$(Y)_{i,t}$ = Financial growth of firm i at time t
$X_{i,t}$ = the value for independent variables of firm i at time t
$\beta_i$ = is the beta coefficient to be determined
$B_0$ = the alpha coefficient representing the constant term
$\varepsilon$ = refers to the error term

The panel data model before including the intervening variable is as presented below: -

$$FG_{i,t} = \beta_0 + \beta_1 STD_{i,t} + \beta_2 LTD_{i,t} + \beta_3 RE_{i,t} + \beta_4 SC_{i,t} + \varepsilon \quad (2)$$

Where;
$FG_{i,t}$ = Financial growth measured by growth in earning per share and growth in market capitalization of firm i at time t
$STD_{i,t}$ = Short-Term Debt of firm i at time t
$LTD_{i,t}$ = Long-Term Debt of firm i at time t
$RE_{i,t}$ = Retained Earning of firm i at time t
$SC_{i,t}$ = Share Capital of firm i at time t
i = Firms listed at NSE from 2008 to 2017
t = Time period (2008-2017)
$\varepsilon$ = Error term

Moreover, the study examined the intervening effect of firm size on the relationship between financial structure and financial growth of non-financial firms listed at the Nairobi securities exchange. The study modified the dynamic panel data model by Baños-Caballero, S., García-Teruel, P. J., and Martínez-Solano, P. as depicted in equation 2 [16]. The Baron, R. M., and Kenny, D. A. approach in testing for mediation was employed. For the intervening effect to be considered positive, four conditions should be fulfilled: One, the independent variable is significantly related to the dependent variable in the absence of the intervening variable [17]. The intervening variable (firm size) interacted with each measure of financial growth.

*Step 1*
The independent variable was significantly related to the dependent variable in the absence of the intervening variable.

$$Y_{i,t} = \beta_0 + \beta_1 X_{i,t} + \varepsilon \quad (3)$$

*Step 2*
The independent variable was significantly related to the intervening variable.

$$Z_{i,t} = \beta_0 + \beta_1 X_{i,t} + \varepsilon \quad (4)$$

*Step 3*
The intervening variable was significantly related to the dependent variable.

$$Y_{i,t} = \beta_0 + \beta_1 Z_{i,t} + \varepsilon \quad (5)$$

*Step 4*
By controlling the effect of the intervening variable on the dependent variable, the effect of the independent variable on the dependent variable was insignificant in the presence of the intervening variable.

$$Y_{it} = \beta_0 + \beta_1 X_{i,t} + \beta_1 Z_{i,t} + \varepsilon \quad (6)$$

Where;
$Y_{i,t}$ = Financial growth
$X_{i,t}$ = Financial structure/Total finance (Composite function)
$Z_{i,t}$ = Firm size (Intervening variable)
i = Firms listed at NSE from 2008 to 2017
t = Time period (2008-2017)

## 4. Results and Discussion

### 4.1. Descriptive Statistics

*Table 1. Descriptive Statistics.*

| Variable | Obs | Mean | Std. Dev. | Min | Max | CV |
|---|---|---|---|---|---|---|
| Short Term Debt | 360 | 0.29146 | 0.255896 | 0.007901 | 2.535623 | 0.87798 |
| Long Term Debt | 360 | 0.200195 | 0.186595 | 0.000000 | 1.126967 | 0.932066 |
| Retained Earnings | 360 | 0.276984 | 0.327572 | -1.60575 | 1.05154 | 1.182639 |
| Share Capital | 360 | 0.100219 | 0.156585 | 0.001601 | 1.139994 | 1.562428 |
| Firm Size in million KES | 360 | 28400.00 | 56200.00 | 40.1960 | 377000.00 | 1.978873 |
| Earnings Per Share | 360 | 6.468265 | 15.03232 | -46.744 | 100.0483 | 2.324011 |
| Market Capitalization in million KES | 360 | 24600.00 | 77300.00 | 116.000 | 721000.00 | 3.142276 |

The descriptive results show that the mean value for short-term debt was 0.29146, with a standard deviation of 0.255896, a minimum of 0.007901, and a maximum of 2.535623. The mean of 29.146% implies that the assets and operations of listed non-financial firms were financed by short-term debt. Long-term debt had a mean of 0.200195, a standard deviation of 0.18695, a minimum of 0.000000, and a maximum of 1.126967. The mean of 20.02% implies that the assets and operations of listed non-financial firms were financed by long-term debt. Retained earnings had a mean of 0.276984, a standard deviation of 0.156585, a minimum of -1.60575, and a maximum of 1.05154. The mean of 27.70% implies that the assets and operations of listed non-financial firms were financed by retained earnings. Further, share capital had a mean of 0.100219, a standard deviation of 0.327572, a minimum of 0.001601, and a maximum of 1.139994. The mean of 10.02% implies that the assets and operations of listed non-financial firms were financed by Share capital.

The findings show that Non-financial firms prefer short-term debt, followed by retained earnings, then long-term debt, and finally share capital. This does not confirm the proposition of the Pecking Order theory. The Pecking Order theory states that firms prioritize their sources of financing from internal financing to equity. Therefore, internal financing is used first and when it depletes, debt is issued and when it is no longer sensible to issue any more debt, equity is issued.

The coefficient of variation (CV) results indicates that the variation was higher. The coefficient of variation shows the extent of variability of data in a sample in relation to the mean of the population. The coefficient of variation allows investors to determine how much volatility, or risk, is assumed in comparison to the amount of return expected from investments.

### 4.2. Correlation Analysis

In order to get an overview of the association between the dependent and independent variables, the researcher conducted a pairwise correlation analysis. The analysis aims at testing for the existence of multicollinearity and it is ideal for eliminating variables that are highly correlated [24]. The study conducted a correlation analysis between financial structure and financial growth measured using growth in earnings per share and growth in market capitalization. Table 2 shows the correlation matrix of short-term debt, long-term debt, retained earnings, and share capital and growth in earnings per share.

*Table 2. Correlation Matrix.*

|  | Growth in EPS | Growth in Market Capitalization | Short Term Debt | Long Term Debt | Retained Earnings | Share Capital |
|---|---|---|---|---|---|---|
| Growth in EPS | 1.000 |  |  |  |  |  |
| Growth in Market Capitalization |  | 1.000 |  |  |  |  |
| Short Term Debt | 0.7108 | 0.7212 | 1.000 |  |  |  |
|  | 0.0001 | 0.000 |  |  |  |  |
| Long Term Debt | 0.6399 | 0.8325 | 0.4185 | 1.000 |  |  |
|  | 0.000 | 0.000 | 0.0245 |  |  |  |
| Retained Earnings | 0.8288 | 0.623 | 0.4929 | 0.2259 | 1.000 |  |
|  | 0.000 | 0.0195 | 0.000 | 0.000 |  |  |
| Share Capital | 0.7495 | 0.755 | 0.3933 | 0.1104 | 0.2072 | 1.000 |
|  | 0.0045 | 0.0032 | 0.000 | 0.0362 | 0.0001 |  |

The correlation results found that short-term debt and growth in earnings per share have a high positive and significant association ($\beta=0.7108$, r=0.001). The results found that long-term debt has a high positive and significant association with growth in earnings per share ($\beta=0.6399$, r=0.000). Long-term debt involves strict contractual covenants between the firm and issuers of debt which is usually associated with high agency and financial distress costs. High long-term debt levels in the firm are not conducive to the effective operations of the firm since they increase the risk of bankruptcy. Retained earnings have a high positive and significant correlation with growth in earnings per share ($\beta=0.8288$, r=0.000). Retained earnings are the most important source of financing growth for the firm. It was also established that share capital has a moderately high positive correlation with growth in earnings per share ($\beta=0.7495$, r=0.0045).

Further, the correlation results found that short-term debt has a high positive correlation with growth in market capitalization ($\beta=0.7212$, r=0.000). There was a high positive correlation between long-term debt and growth in market capitalization ($\beta=0.8325$, r=0.000). There was a high positive correlation between retained earnings and growth in market capitalization ($\beta=0.623$, r=0.0195). The study showed there was a high positive correlation between share capital and growth in market capitalization ($\beta=0.755$, r=0.0032). Share capital is an indication of value contributed to a firm by its shareholders at some time in the past.

### 4.3. Panel Regression Analysis

The panel regression analysis was conducted between financial structure (short-term debt, long-term debt, retained earnings, and share capital), Firm size and financial growth (measured using both growth in earnings per share and growth in market capitalization).

#### 4.3.1. Financial Structure, Firm Size, and Financial Growth (Measured by Growth in Earnings Per Share)

The random effect model was estimated between financial structure, Firm size, and financial growth (measured by growth in earnings per share). The panel regression on the financial structure, firm size, and financial growth measured by growth in earnings per share is shown in Table 3.

*Table 3. Financial Structure, Firm Size, and Financial Growth (Measured by Growth in Earnings Per Share).*

| **Direct Effect** |  | **Intervening Effect** |  |  |  |
|---|---|---|---|---|---|
|  | **Model I** | **Model 2 (Step 1)** | **Model 2 (Step 2)** | **Model 2 (Step 3)** | **Model 2 (Step 4)** |
| Short Term Debt | 0.024095** (0.013) |  |  |  |  |
| Long Term Debt | 0.864088** (0.000) |  |  |  |  |
| Retained Earnings | 0.951749** (0.015) |  |  |  |  |
| Share Capital | 0.007016** (0.000) |  |  |  |  |
| Composite of financial structure (Total finance) |  | 1.229971** (0.013) | 0.178965** (0.000) |  | 1.35455 (0.141) |
| Firm Size |  |  |  | 1.13874** (0.036) | 1.6834 (0.424) |
| $\beta_0$ | 2.254497 (0.068)* | -20.3387 (0.321) | 5.854919** (0.000) | -.79045 (0.37) | -6.48209 (0.809) |
| R-squared: | 0.6136 | 0.6009 | 0.6034 | 0.6128 | 0.6305 |
| Wald chi2 (4) | 68.43** | 5.6800** | 77.40** | 4.2900** | 2.47 |
| Prob > chi2 | 0.0000 | 0.0363 | 0.000 | 0.002 | 0.2914 |

P-values in parentheses where *** <0.1, ** <0.5 and *<0.05

Model 1 depicts the results between financial structure and financial growth measured by growth in EPS. The results show the coefficient of determination R-square of 0.6136. This means that short-term debt, long-term debt, retained earnings and share capital explain 61.36% of variations in financial growth as measured by the growth in earnings per share. Further, the results revealed that short-term debt has a positive and significant effect on financial growth as measured by growth in earnings per share ($\beta=0.024095$, $p=0.013$). This implies that a unit increase in short-term debt results in an increase in the growth in earnings per share by 0.024095 units. Long-term debt has a positive and significant effect on financial growth as measured by growth in earnings per share ($\beta=0.864088$, $p=0.000$). This implies that a unit increase in long-term debt results in an increase in the growth in earnings per share by 0.864088 units. The results agree with Mwangi, L. W., Muathe, S. M. A., and Kosimbei, G. K. who concluded that the majority of firms at NSE use long-term debt to finance their assets [57]. Further, the results revealed that retained earnings has a positive and significant effect on financial growth as measured by growth in earnings per share ($\beta=0.951749$, $p=0.015$). This implies that a unit increase in retained earnings results in an increase in the growth in earnings per share by 0.951749 units. Lastly, share capital has a positive and significant effect on financial growth as measured by earnings per share ($\beta=0.007016$, $p=0.000$). Share capital is an indication of value contributed to a firm by its shareholders at some time in the past.

Model 2 depicts the results of the intervening effect of the firm size on the effect of financial structure on the financial growth measured by growth in earnings per share. In step one, the influence of the composite of financial structure/total finance on financial growth as measured by growth in earnings per share is significant ($\beta=1.229971$, $R^2=0.6009$, $p<0.05$), thus satisfying the first condition which states that the independent variable is significantly related to the dependent variable in the absence of the intervening variable.

The second step involved the regression of the composite of financial structure/total finance on the firm size. The results presented in Table 3 show that the influence of the composite of financial structure/total finance on the firm size is significant ($\beta=0.178965$, $R^2=0.6034$, $p<0.05$), thus satisfying the second condition which states that the independent variable should be significantly related to the intervening variable, for the process to continue to step 3.

The third step was intended to test for the influence of the firm size on growth in earnings per share. As shown in Table 3, the influence of the firm size on growth in earnings per share was significant ($\beta=1.13874$, $R^2=0.6128$, $p<0.05$), thus satisfying the third condition that the intervening variable should be significantly related to the dependent variable, for the process of testing for intervention to continue to the final step. The firm size is significantly related to the growth in earnings per share.

The fourth step was to test if the firm size and composite of financial structure/total finance is insignificantly related to growth in earnings per share. Here, the firm size and financial structure were tested to see if it predicts financial growth as measured by growth in earnings per share. Table 3 shows that the influence of the total finance/composite of financial structure on growth in earnings per share was insignificant in the presence of the firm size ($\beta=1.35455$, $R^2=0.6305$, $p>0.05$), and thus satisfying the fourth condition which states that the effect of the total finance/composite of financial structure on growth in earning per share is insignificant in the presence of the firm size. This implies that the firm size accounts for growth in earnings per share. The results agree with Anafo, S. A., Amponteng, E., and Yin, L. that firm size has an intervening effect on the relationship between financial structure and earnings per share [9]. Firm size has the potential to influence financial structure [88]. The results also agree with Yuliza, A. that firm size intervenes in the relationship between earnings per share and stock prices [90]. Moreover, Pouraghajan, A., Mansourinia, E., Bagheri, S. M. B., Emamgholipour, M., and Emamgholipour, B. noted that there is a positive and significant relationship between financial ratios and firm size with earnings per share [76].

### 4.3.2. Financial Structure, Firm Size, and Financial Growth (Measured by Growth in Market Capitalization)

*Table 4. Financial Structure, Firm Size and Financial Growth (Measured by Growth in Market Capitalization).*

|  | Direct Effect | Intervening Effect | | | |
|---|---|---|---|---|---|
|  | Model 1 | Model 2 (Step 1) | Model 2 (Step 2) | Model 2 (Step 3) | Model 2 (Step 4) |
| Short Term Debt | 0.028529** (0.006) | | | | |
| Long Term Debt | 0.958656** (0.000) | | | | |
| Retained Earnings | 0.43784** (0.004) | | | | |
| Share Capital | 0.09635** (0.001) | | | | |
| Composite of financial structure (Total Finance) | | -0.16209 (0.92) | 0.178965 (0.547) | | 0.04852** (0.044) |
| Firm Size | | | | 2.57561 (0.478) | 2.56464 (0.483) |
| $\beta_0$ | 6.050879** (0.000) | 4.5177 (0.901) | 5.854919 (0.000) | 26.364 (0.465) | 27.35388 (0.576) |
| R-squared: | 0.6557 | 0.615 | 0.6134 | 0.614 | 0.6469 |
| Wald chi2 (4) | 59.57** | 0.0100 | 1.40 | 0.4780 | 19.22 |
| Prob > chi2 | 0.0000 | 0.9197 | 0.9198 | 0.0500 | 0.0319 |

P-values in parentheses where *** <0.1, ** <0.5 and * <0.05

The random effect model was estimated between financial structure, Firm size, and financial growth (measured by growth in market capitalization). The panel regression on the financial structure, firm size, and financial growth measured by growth in market capitalization is shown in Table 4.

Model 1 depicts the results between financial structure and financial growth measured by growth in market capitalization. The results show the coefficient of determination R-square of 0.6557 which indicates short-term debt, long-term debt, retained earnings and share capital account for 65.57% of the variation in financial growth as measured by growth in market capitalization. Further, the results revealed that short-term debt has a positive and significant effect on financial growth as measured by growth in market capitalization ($\beta=0.028529$, $p=0.006$). The results imply that a unit increase of short-term debt leads to growth in market capitalization by 0.028529 units. Long-term debt has a positive and significant effect on financial growth as measured by growth in market capitalization ($\beta=0.958656$, $p=0.000$). This result implies that a unit increase of long-term debt leads to growth in market capitalization by 0.958656 units. The results agree with Lixin, X., & Lin, C. in a study on the relationship between debt financing and the market value of listed real estate firms in China that, long-term borrowing and commercial credit financing have a positive correlation with the firm's market value [48]. Retained earnings had a positive and significant effect on financial growth as measured by growth in market capitalization ($\beta=0.43784$, $p=0.004$). This implies that a unit increase in retained earnings results in an increase in the growth in earnings per share by 0.43784 units. Lastly, share capital has a positive and significant effect on financial growth as measured by growth in market capitalization ($\beta=0.09635$, $p=0.001$). The results agree with Ebaid, E. I. share capital has a significant relationship with return on assets but not with return on equity [26].

Model 2 depicts the results of the intervening effect of the firm size on the effect of financial structure on the financial growth measured by growth in market capitalization. In step one, the influence of the composite of financial structure/total finance on financial growth as measured by growth in market capitalization is insignificant ($\beta=-0.16209$, $R^2=0.615$, $p>0.05$), thus not satisfying the first condition which states that the independent variable is significantly related to the dependent variable in the absence of the intervening variable.

The second step involved regression of total finance/composite of financial structure on the firm size. The results presented in Table 4 show that the influence of total finance/composite of financial structure on the firm size is insignificant ($\beta=0.178965$, $R^2=0.6134$, $p>0.05$), thus not satisfying the second condition which states that total finance is significantly related to the firm size, for the process to continue to step 3.

The third step was intended to test for the influence of the firm size on growth in market capitalization. As shown in Table 4, the influence of firm size on growth in market capitalization was insignificant ($\beta=2.57561$, $R^2=0.614$, $p>0.05$), thus not satisfying the third condition that the intervening variable is significantly related to the dependent variable for the process of testing for mediation to continue to the final step. The Firm size is insignificantly related to the growth in market capitalization.

The fourth step was to test if the firm size and composite financial structure are significantly related to the growth in market capitalization. Here, the firm size and total finance/composite of the financial structure was tested to see if it predicts financial growth as measured by growth in market capitalization. Table 4 shows that the influence of the total finance/composite of financial structure on growth in market capitalization was significant in the presence of the firm size ($\beta=0.04852$, $R^2=0.6469$, $p<0.05$), and thus not satisfying the fourth condition which states that, the effect of total finance/composite of financial structure on growth in market capitalization is insignificant in the presence of the firm size. This implies that the firm size does not account for growth in market capitalization.

### 4.4. Summary of Hypotheses

*Table 5. Summary of Hypotheses.*

| Objective | Hypothesis | Rule | P-value | Comment |
|---|---|---|---|---|
| To establish the effect of Short-term debt on the financial growth of non-financial firms listed at NSE. | $H_{o1}$: There is no significant effect of Short-term debt on the financial growth of non-financial firms listed at NSE. | Reject Ho if p-value <0.05 | p<0.05 | The results reject the hypothesis; therefore, there is a significant effect of Short-term debt on the financial growth of non-financial firms listed at NSE measured using both growth in EPS and growth in market capitalization. |
| To assess the effect of Long-term debt on the financial growth of non-financial firms listed at NSE. | $H_{o2}$: There is no significant effect of Long-term debt on the financial growth of non-financial firms listed at NSE. | Reject Ho if p-value <0.05 | p<0.05 | The results reject the hypothesis; therefore, there is a significant effect of Long-term debt on the financial growth of non-financial firms listed at NSE measured using both growth in EPS and growth in market capitalization. |
| To determine the effect of Retained earnings on the financial growth of non-financial firms listed at NSE. | $H_{o3}$: There is no significant effect of Retained earnings on the financial growth of non-financial firms listed at NSE. | Reject Ho if p-value <0.05 | p<0.05 | The results reject the hypothesis; therefore, there is a significant effect of retained earnings on the financial growth of non-financial firms listed at NSE measured using both growth in EPS and growth in market capitalization. |

| Objective | Hypothesis | Rule | P-value | Comment |
|---|---|---|---|---|
| To examine the effect of share capital on the financial growth of non-financial firms listed at NSE. | H$_{o4}$: There is no significant effect of share capital on the financial growth of non-financial firms listed at NSE. | Reject Ho if p-value <0.05 | p<0.05 | The results reject the hypothesis; therefore, there is a significant effect of share capital on the financial growth of non-financial firms listed at NSE measured using both growth in EPS and growth in market capitalization. |
| To explore the intervening effect of firm size on the effect of financial structure on the financial growth of non-financial firms listed at NSE. | H$_{o5}$: Firm size does not significantly intervene in the effect of financial structure on the financial growth of non-financial firms listed at NSE. | Four steps of testing the intervening effect must be fulfilled | The four steps are fulfilled with growth in EPS as a measure of financial growth. However, the four steps were not fulfilled with growth in market capitalization as a measure of financial growth. | Firm size significantly intervenes in the effect of financial structure on the financial growth of non-financial firms listed at NSE measured using growth in earnings per share. Firm size does not significantly intervene in the effect of financial structure on the financial growth of non-financial firms listed at NSE measured using growth in market capitalization. |

# 5. Conclusion and Recommendations

*5.1. Conclusion*

The study concluded that short-term debt has a positive and significant relationship with financial growth measured using both growth in earnings per share and growth in market capitalization. Earnings per share is considered as an important accounting indicator of risk, entity performance, and corporate success. It is used to forecast potential growth in future share prices because changes in EPS are often reflected in share price behavior. EPS is a useful investment decision tool for investors because it indicates future prospects and growth. Short-term assets should be financed with short-term liabilities and long-term assets should be financed with long-term liabilities. Short-term financing is primarily concerned with the analysis of decisions that affect current assets and current liabilities.

It was further concluded that long-term debt has a positive and significant relationship with financial growth using both growth in earnings per share and growth in market capitalization. Long-term debt is the most preferable source of debt financing among well-established corporate institutions, mostly by virtue of their asset base and collateral is a requirement by many deposits-taking financial institutions. Long-term debt involves strict contractual covenants between the firm and issuers of debt which is usually associated with high agency and financial distress costs. A high long-term debt level is not conducive to the effective operations of the firm since it increases the risk of bankruptcy.

It was also concluded that retained earnings has a positive and significant relationship with financial growth using both growth in earnings per share and growth in market capitalization. The amount of retained earnings is an important issue to investors and other stakeholders because it is another way to evaluate the effectiveness of management to bring improvement in the market value of the firm. Retained earnings is important because it has a significant effect on a firm's stock prices. In making their decisions, investors mostly look for firms with a high return on retained earnings that is reinvested regularly.

The study concluded that share capital has a positive and significant relationship with financial growth measured using growth in earnings per share. Share capital is the total capital of a company divided into shares. A joint stock company should have capital in order to finance its activities. Share capital is an indication of value contributed to a firm by its shareholders at some time in the past. It was further concluded that share capital has a positive and significant relationship with financial growth measured using growth in market capitalization.

The study further concluded that firm size has an intervening effect on financial structure and financial growth measured using growth in earnings per share. However, firm size was observed to have no intervening effect on the relationship between financial structure and financial growth measured using growth in market capitalization. Large firms have more resources and capacity to undertake more product lines and higher production capacity together with organizational resources. This enables large firms to improve their financial performance since they can easily mitigate risks as compared to small firms.

*5.2. Recommendations*

Based on the findings, the study recommends that the management of non-financial firms listed at the Nairobi Securities Exchange need to balance between financing options. The choice depends upon which source of funding is most accessible for the firm. The short-term debt should be used to keep the business running during times when the revenue stream temporarily is insufficient to meet operational needs. In addition, short-term financing should be aligned with a firm's operational needs. A firm may find itself in a crisis if they are unable to renew its debt usually because of some negative news, real or otherwise. Most failures of institutions can be due to the unavailability of short-term funding.

In addition, the study recommends that firms looking for long-term financing can go for equity or preference shares and debentures. Long-term debt is not conducive to financing activities in the process of generating profits. High levels of long-term debt increase the number of interest payments that are expected to be paid regularly, thus lowering the profitability of the companies. The firms should only use long-term debts if other financing options are not available such as short-term debts. Long-term debts are the most preferable sources of debt financing among well-established corporate institutions mostly by virtue of their asset base.

The study recommends that the management of non-financial firms listed at NSE need to encourage its shareholders to re-invest back their earnings rather than consuming them as dividends. It was noted that retained earnings significantly influence financial growth measured by growth in earnings per share. Notably, retained earnings are a sacrifice made by equity shareholders. As an internal source, retained earnings are readily available for use. Also, retentions are cheaper than external equity, do not cause ownership dilution, and have got a positive connotation as the stakeholders perceive that the firm has potential investment opportunities. Since only a few firm financing options are available, firms prefer to retain more earnings and plow them back into operations especially when they have viable investment opportunities.

The study recommends that share capital should be used as a last resort. Even though share capita has a positive effect on financial growth, it should only be preferred only when the financing option through short-term debts has reached optimal and the company lacks an alternative method to finance its activities. Equity financing should be important to any firm if the proceeds are used as a way of raising capital for major expansions, asset growth, or acquisitions which may require heavy funding. A joint stock company should have capital in order to finance its activities. Increasing a company's share capital can lead to the shares of existing shareholders becoming diluted.

Moreover, the study recommends that the firms need to increase their firm size and this can be achieved by increasing the number of branches. There is a need to increase the size by increasing various aspects of the customer base, net assets, deposit liabilities, and market share. Lenders and other potential investors are much more confident to lend out funds to companies that have a larger market share. Larger firms obtain benefits from their size and diversification because they can borrow with lower costs and survive economic disasters with more resilience than smaller firms and thus generate more profit. Large firms have more resources and capacity to undertake more product lines and higher production capacity together with organizational resources. This enables large firms to improve their financial growth since they can easily mitigate risks as compared to small firms.